\documentclass[intlimits,twoside,a4paper]{article}

\usepackage{amsmath,amssymb}
\usepackage{graphicx}

\usepackage[T2A]{fontenc}
\usepackage[cp1251]{inputenc}

\usepackage[eqsecnum]{cmpj3}
\issue{2020}{23}{3}{33301}
\doinumber{10.5488/CMP.23.33301}
\title[Investigation on quality of mango fruit]%
{Detection of odor quality and ripening stage of Mangifera indica L. by graphdiyne nanosheet --- a DFT outlook}%
\author[V. Nagarajan, R. Chandiramouli]{V. Nagarajan, R. Chandiramouli}
\address{
School of Electrical and Electronics Engineering, Shanmugha Arts Science Technology and Research Academy (SASTRA) Deemed University, Tirumalaisamudram, Thanjavur, Tamil nadu --- 613 401, India 
}
\date{Received December  28, 2019, in final form March 21, 2020}

\begin{document}

\maketitle

\begin{abstract}

Using first-principles calculation, geometrical stability together with electronic properties of graphdiyne nanosh-
\noindent eet (Gdn-NS) is investigated. The structural stability of Gdn-NS is established with the support of phonon band structure and cohesive energy. The main objective of the present study is to check the odor quality of Mangifera indica L. (mangoes) fruits during the various ripening stage with the influence of Gdn-NS material. In addition, the adsorption of various volatiles, namely ethyl butanoate, myrcene, (E,Z,Z)-1,3,4,8-undecatetraene and $\gamma$-octalactone aromas on Gdn-NS is explored with the significant parameters including Bader charge transfer, energy gap, average energy gap changes and adsorption energy. The sensitivity of volatiles emitting from various ripening stages of mango on Gdn-NS were explored with the influence of density of states spectrum. The outcomes of the proposed work help us to check the ripening stage and odor quality of Mangifera indica L. by Gdn-NS material using density functional theory.

\end{abstract}

\section{Introduction}

Mango (Mangifera indica L) is one of the most significant tropical fruits, owing to its pleasant flavor and taste. The cultivation of mangoes  started four thousand years back in eastern India. Nowadays, mango trees are grown all over the tropical and certain subtropical zones of the world~\cite{1}. The main aspect contributing to the attractiveness of mango is its marvelous flavor. Especially in western countries, mango fruits are generally considered with regard to high quality and low fiber content, a balanced acidity, sweet taste and an aroma-profile low in terpeny and high in fruity notes. Moreover, depending upon their shape, color and size, the properties of  the flavour of mangoes are strongly affected by the cultivar \cite{1}.
			
Many research works have been conducted based on volatile organic compounds (VOCs) of mangoes in various cultivars \cite{2}. This leads to the identification of a few hundred of compounds to date \cite{2}. Importantly, it revealed that the maturity \cite{1} and the geographic origin \cite{3} highly influence the volatile-profile of tasty fresh mangoes. The potential application on aroma-extract-dilution-analysis (AEDA) \cite{4} towards headspace extracts was observed from mangoes of various cultivars grownup in Brazil. Moreover, the ethyl butanoate molecules have a fruity-smell, which is one of significant aroma-active volatiles in all samples studied, whereas the other important volatiles, namely, $\alpha$-pinene, $\delta$-3-carene and 2- and 3-methylbutanoates are significant depending upon the cultivators \cite{5}. A detailed investigation including the detection of 20 mango varieties monitored by odor activity values (OAV) calculation leads to the identification of ethyl butanoate, methyl benzoate, decanal, (E,Z)-2,6-nonadienal which are potentially most significant mango aroma volatiles \cite{6}.
			
			During the ripening period of mangoes, we observe the changes of the size, color, shape, firmness including the aroma of the mango fruit. In addition to these parameters, pulp composition, acidity, soluble solid content, moisture content and physiological weight, can be utilized to measure the maturity and quality of the mango fruit \cite{7}. Analyzing the chemical properties to measure the fruit ripening is most significant and advantageous owing to its direct link with the consuming quality. Nevertheless, previous techniques were often invasive, demanding the mangoes to be pulped at the earlier chemical analysis. Nowadays, non-destructive techniques are used to measure the quality of the mango fruit based on electronic noses \cite{8} and near-infrared spectroscopy \cite{9}. Lebrun et al. \cite{10}  proposed an electronic nose to assess the variations between the aromas of Keitt, Kent, Cogshall mango cultivars. Jha et al. \cite{7} have also reported these mangoes and probing the quality of fruit using non-destructive detectors.

		Many analytical methods have been utilized to assess volatiles from mango headspace including gas-chromatography (GC), mass-spectrometry (MS), proton-transfer reaction (PTR) MS and chemical-ionization reaction (CIR) MS \cite{11,12}. Moreover, we learned from the literature survey that mango fruits are the most important tropical fruits attracted by many vertebrates and of course by human beings. Further, it had a wide range of acceptance due to their flavor value in which the aroma of mango fruit plays a significant role to determine the odor quality of mangoes.

		Carbon is one of the most abundant species that form various kinds of allotropes. It exhibits three dissimilar hybridized states including $sp^3$, $sp^2$ and $sp$. Further, carbon species adopt various bonding states and are proficient in bond with other elements or with themselves. The generally known allotropes of carbon are graphene \cite{13}, fullerenes \cite{14}, carbon nanotubes \cite{15} and nanobuds \cite{16}. Moreover, the preparation of new carbon allotropes along with high stability, novel bonding characteristics, potential applications, and unique features are the present interest among materials scientists \cite{17}. Besides, the new form of non-natural carbon allotropes includes graphite/graphene. Graphyne including graphdiyne is in the main focus among researchers owing to its intriguing mechanical, optical and electronic characteristics with similar structures \cite{18}. Besides, these carbon allotropes are found to be a good candidate for potential applications in nanoelectronics including energy storage \cite{19}. Furthermore, graphdiyne has one additional acetylenic linkage when related to graphyne, which is different in all separate unit cells. Comparatively, graphdiyne material is softer than graphyne and graphene for a thickness magnitude of 0.320 nm \cite{20}. Pei et al. \cite{21}  found that the calculated energy band gap of graphdiyne ranges from 0.47 eV to 1.12~eV. Haley et al. \cite{22} anticipated graphdiyne material as far back as in 1997 using the ab initio method.

		Experimentally, various methods are utilized to prepare graphdiyne on different substrates \cite{23,24}. Interestingly, graphdiyne was explored in many applications such as photocatalysis \cite{25}, spintronics~\cite{26}, metal-free catalysis \cite{27} and transistors \cite{28}. Moreover, graphdiyne nanosheet is a promising material for the diffusion of CH$_4$, CO and H$_2$ molecules which is explored and confirmed by first-principles calculations~\cite{29}. Chen et al. \cite{30}  investigated the sensing behavior of amino acid molecules on graphdiyne material. In our previous work, we explored the interaction behavior of NH$_3$, trimethyl amine (TMA) and dimethyl amine (DMA) on Gdn-NS using density functional theory (DFT) technique \cite{31,32}. As we studied the sensing properties of graphdiyne nanosheet towards various vapor, we thus decided to explore the odor quality of mangoes through the interaction behavior of four different vapor molecules namely, (1) ethyl butanoate (fruity), (2) myrcene (terpeny), (3) (E,Z,Z)-1,3,4,8-undecatetraene (pineapple) and (4) $\gamma$-octalactone (coconut)  emitted from Indian mango fruits (Alfonso, Royal special) on graphdiyne nanosheet (Gdn-NS). In this work, electronic properties, geometrical stability and interaction behavior of aforementioned mango volatile vapors on Gdn-NS are studied for the very first time, which can be used to check the odor quality of mangoes at the ripening stage.

\section {Computational methods}

We involved a suitable package SIESTA \cite{33} for optimizing the zigzag Gdn-NS and investigated the electronic characteristics of Gdn base material.  In this work, we facilitated GGA/PBE exchange-correlation functional including van der Waals-density functional (vdW-DF) dispersion correction and carried out local structural optimization \cite{34}. The vdW-DF helps to eliminate the underestimation of the band gap. In addition, the PBE functional is an upright choice for exploring the electronic properties of Gdn-NS, which is validated with published work of Guangfu Luo et al. \cite{35}. Besides, the Troullier-Martin (TM) pseudopotentials have been facilitated to study the core-valence interaction \cite{36}. The grid mesh cutoff energy is fixed to 600 eV and we carried out the complete structural relaxation with conjugate gradient (CG) along with prominent double-zeta polarization (DZP) basis set \cite{37,38} until the Hellmann-Feynman force was obtained to 0.02 eV/\AA. Indeed, the charge density, band gap, electron localization and density of states (DOS) spectrum of Gdn-NS material were calculated using SIESTA code. The Brillouin zone (BZ) integration of Gdn-NS was simulated with 5 x 5 x 1 Monkhorst-Pack $k$-point mesh \cite{39}. Besides, the vacuum space of 16 $\AA$ along the perpendicular direction of Gdn nanosheet is utilized to eliminate the interlayer interaction. Transmission of electrons is studied with the support of Bader-atoms-in-molecules (BAIM) analysis \cite{40}. In this work, the interaction behavior of ethyl butanoate, myrcene, (E,Z,Z)-1,3,4,8-undecatetraene and  $\gamma$-octalactone were explored with the support of SIESTA package. In the proposed study, we  also performed the quantum Monte Carlo (QMC) simulation utilizing QMCPACK utility package \cite{41} to explore ground-level energetics of Gdn-NS, mainly for cohesive energy $(E_{\text{coh}})$ and DOS spectrum. This implies that the cohesive formation energy of Gdn base material is overvalued in DFT technique when compared to the QMC simulated energies, as suggested by previously reported studies \cite{42,43}.
	The adsorption energy of mango volatiles on Gdn-NS can be found using equation (\ref{eq1})
	
	\begin{equation}
E_{\text{ad}} = \left[ E(\text{Gdn-{NS}}+\text{VOC})-E(\text{Gdn-NS})-E(\text{VOC})+E(\text{BSSE})\right] 
\label{eq1}
\end{equation}
where $E$(Gdn-NS$+$VOC) illustrates the energy of Gdn-NS$+$VOC complex. $E$(VOC) illustrates the corresponding isolated-energy of ethyl butanoate, myrcene, (E,Z,Z)-1,3,4,8-undecatetraene and $\gamma$-octalactone. In the present study, counterpoise method should be employed to eliminate the effect of overlapping on basis set functions utilizing basis-set-superposition-error (BSSE) \cite{44,45,46}.

\section {Results and discussion}

\subsection {Structural optimization of Gdn-nanosheet}

\begin{figure}[!b]
\centerline{\includegraphics[width=0.65\textwidth]{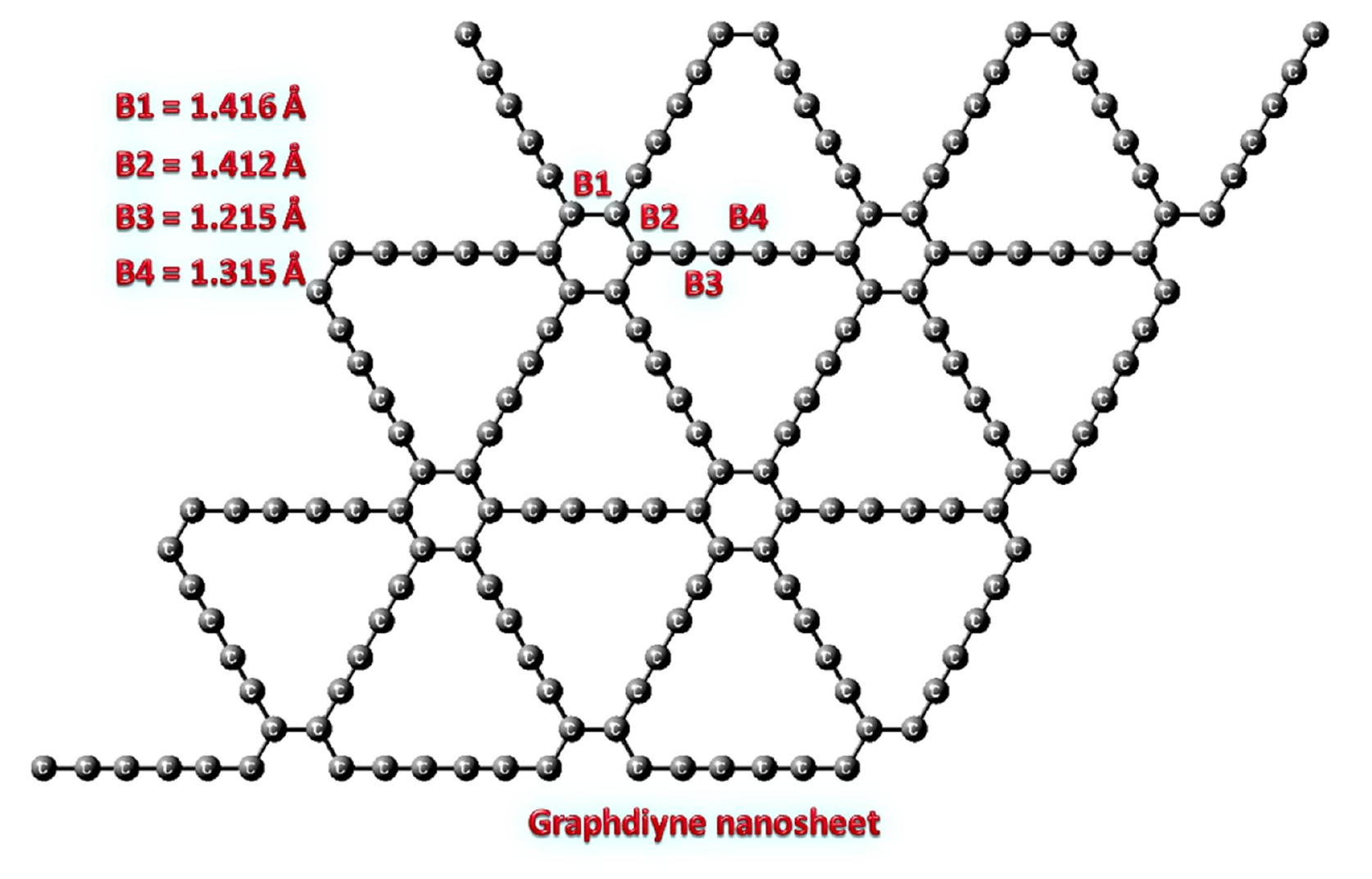}}
\caption{(Colour online) Optimized structure of graphdiyne nanosheet (Gdn-NS).} \label{fig-s1}
\end{figure}

The completely optimized 3 x 3 supercell size of zigzag Gdn-NS system is displayed in figure~\ref{fig-s1}. Clearly, the obtained values of the lattice constants  for an optimized unit cell $a$ and $b$ is 9.492 \AA, which is confirmed with previously reported work \cite{47}. The bond distance between carbon-atoms (BL1, BL2, BL3 \& BL4) on Gdn-NS is presented in figure~\ref{fig-s1}.

The calculated bond length among carbon species in the hexagonal ring is noticed to be 1.416 \AA. Further, the $sp-sp$ hybridization bond distance in acetylenic links is found to be 1.215 \AA. Further, the bond distance between $sp^2$ – $sp^2$ hybridization differs for each pair of C-atoms as presented in figure~\ref{fig-s1}. The lattice constants in the present study are authenticated with the proposed work \cite{47}. The prime objective of this proposed study is to explore geometric stability, and electronic properties of Gdn-NS are calculated through the cohesive energy, band gap, electron density and DOS-spectrum.

\subsection {Geometric stability and electronic properties of Gdn-nanosheet}

\begin{figure}[!b]
\centerline{\includegraphics[width=0.85\textwidth]{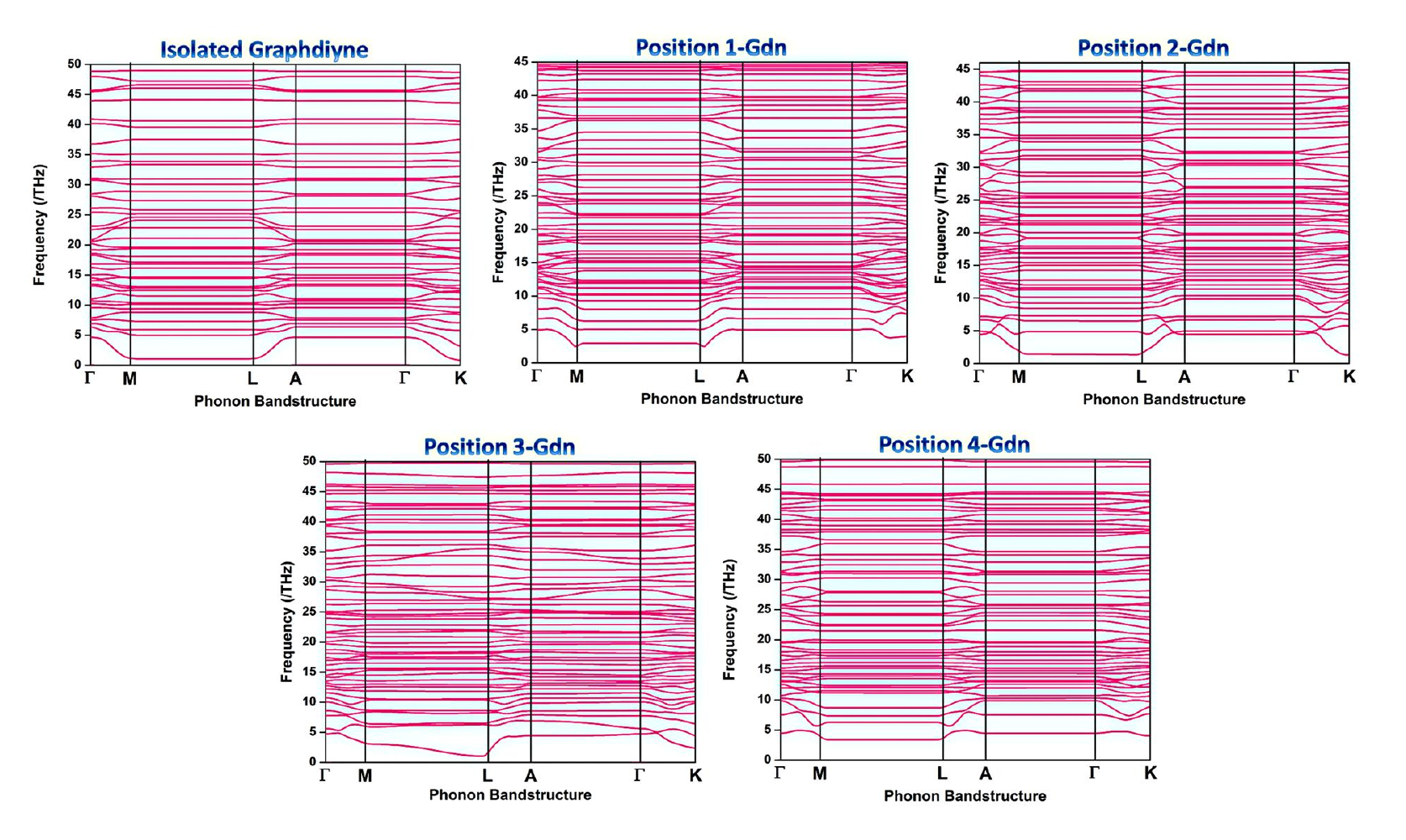}}
\caption{(Colour online) Phonon band structure of isolated Gdn-NS along with the adsorbed mango volatiles.} \label{fig-s2}
\end{figure}

\begin{figure}[!b]
\centerline{\includegraphics[width=0.9\textwidth]{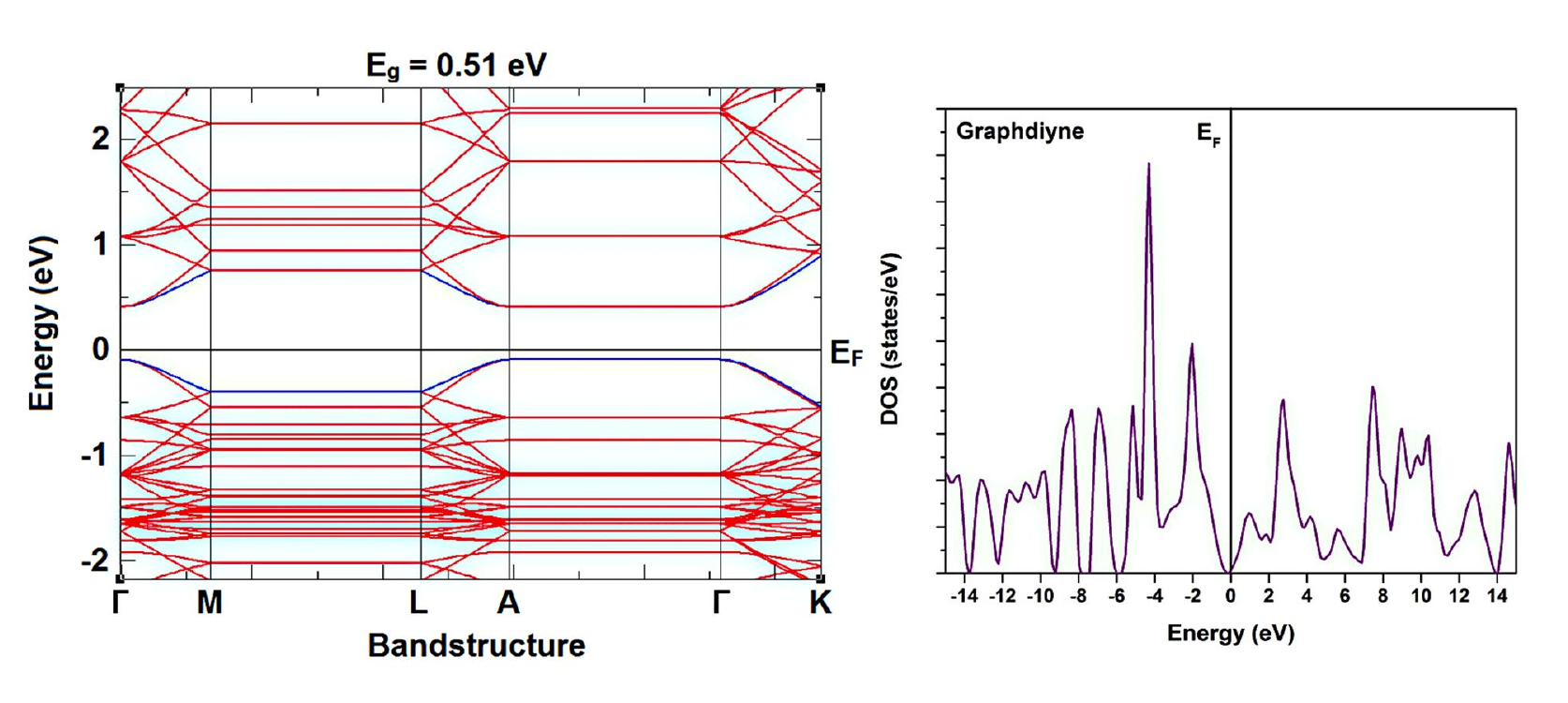}}
\caption{(Colour online) The energy band gap structure and DOS-spectrum of Gdn-NS.} \label{fig-s3}
\end{figure}

The geometrical stability of pristine Gdn-NS was investigated in terms of cohesive energy (E$_\text{coh}$)~\cite{48,49} as shown in equation (\ref{eq3.1}) below

\begin{equation}
	E_{\text{coh}} = (1/n)[E (\text{Gdn-NS})- n E(\text{C})], 
	\label{eq3.1}
\end{equation}
where $E$(C) refers to the individual C-atom energy, and $n$ attribute the whole number of carbon elements in Gdn-NS material. $E$(Gdn-NS) refers to the energy of graphdiyne base material on the above equation. The calculated $E_{\text{coh}}$ of pristine Gdn-NS is obtained to be $-7.796$ eV/atom for PBE functional. In order to obtain more accurate results, we  calculated the $E_{\text{coh}}$ of pristine Gdn-NS using a different hybrid exchange-correlation functional, which includes BLYP, revPBE and RPBE. In that order, the $E_{\text{coh}}$ is obtained to be $-7.083$, $-7.605$ and $-7.604$ eV/atom. Comparatively, BLYP hybrid functional highly influences the $E_{\text{coh}}$ rather than the other exchange-correlation functional. In the present work, we concentrated to study the adsorption behavior of volatiles emitting from mango fruit to estimate its ripening stage and to check its quality. Moreover, GGA-PBE exchange-correlation functional is an optimum choice for studying the interaction behavior of different volatiles on Gdn-NS. In addition, $-7.142$ eV/atom is recorded based on QMC calculation, which is also validated with the proposed work ($-7.20$ eV/atom) \cite{50}. Nevertheless, the $E_{\text{coh}}$ of Gdn-NS in this work is somewhat smaller than graphene nanosheet (7.464~eV) \cite{49}. The magnitude of $E_{\text{coh}}$ of Gdn-NS confirms that the system is stable and  can be used for odor quality checking of mango fruit. Further, to verify the geometrical stability of Gdn-NS, phonon-band structure ($P_{\text{band}}$) is considered. The ($P_{\text{band}}$) of isolated Gdn-NS along with various adsorption sites of mango volatiles is explored. It is revealed that for all possible interaction sites, the ($P_{\text{band}}$) is observed only on positive frequency. Thus, the result of ($P_{\text{band}}$) validates that the Gdn-NS system is noticeably stable and confirms the dynamic stability from the phonon-band structure spectrum for different interaction sites. The ($P_{\text{band}}$) of isolated Gdn-NS with adsorbed mango volatiles is displayed in figure~\ref{fig-s2}. The electronic characteristics of Gdn-NS can be explored in terms of the energy band gap $(E_g)$ structure and DOS-spectrum~\cite{51,52}. The pictorial representation of the band gap structure and DOS-spectrum of pristine Gdn-NS is shown in figure~\ref{fig-s3}. The $E_g$ value of isolated Gdn-NS is calculated to be 0.51 eV over the $\Gamma$-point. The observed $E_g$ value of Gdn-NS in the present study is confirmed with the already mentioned work (0.46 eV \cite{47}).

Luo and co-workers \cite{35} demonstrated the optical, electronic and structural characteristics of Gdn nanostructure and reported the band gap value around 0.52 eV utilizing vdW-DFT methods. Pei and co-workers \cite{21} have observed the energy gap of Gdn material from 0.47 eV to 1.12 eV. The band gap of Gdn base material can be fine-tuned using chemical functionalization, edge border effect and varying the width of Gdn nanosheet as we learned from previous reports. In the present study, we used GGA/PBE functional including vdW-DF dispersion correction and computed the band gap of Gdn-NS. In addition, the underestimation of $E_g$ is noticeably slashed using vdW-DF.
	The localized-electronic-states (LES)	 on Gdn-NS influence the DOS-spectrum in distinct energy intervals. In this pictorial representation (figure~\ref{fig-s3}), the peak maximum ($P_{\text{MAX}}$) is detected much nearer to the Fermi-level-energy (EF) of pristine Gdn-NS, which reveals that it can be used as chemi-resistive based sensor. Further, this peak maximum in distinct energy intervals is perceived due to the bonding character of C-elements mainly in acetylenic linkages of Gdn nanosheet. Usually, the current passing through the chemi-resistive based low dimensional molecular nanodevice is directly proportional to the amount of volatiles evolved from the mango fruit. In addition, the transmission of electrons through the mango volatiles and Gdn-NS material leads to the deviation of the electrical resistance in the base material. Therefore, the DOS-spectrum and the energy gap of pristine Gdn-NS strongly validates that the Gdn-NS can be used as a substrate for chemi-resistive based sensor application.

\subsection {Interaction behavior of mango fruit volatiles on Gdn-nanosheet}

\begin{figure}[!b]
\centerline{\includegraphics[width=0.63\textwidth]{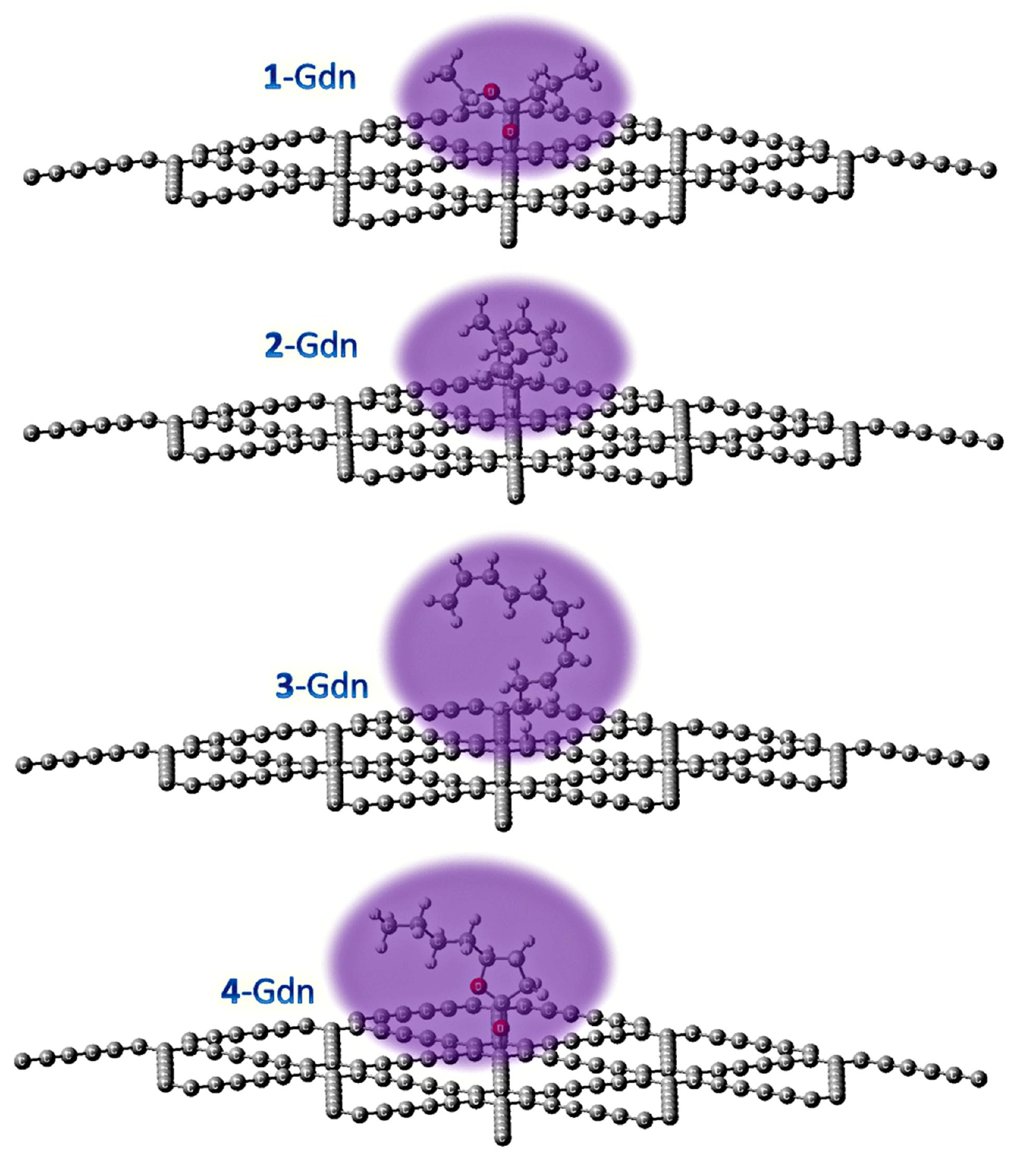}}
\caption{(Colour online) Interaction of Mangifera indica L. volatiles on Gdn-NS-position 1-Gdn, 2-Gdn, 3-Gdn \& 4-Gdn. } \label{fig-s4}
\end{figure}

We have already noticed that mango is one amongst the most popular and significant tropical fruits, especially owing to the delicious flavor and taste. Volatiles that give an aroma of processed and fresh mangoes have been widely explored in many cultivars. Around 270 VOCs have been detected in mango fruit in the past century \cite{53}. Sesquiterpene and monoterpene were noticed to be copious mango volatiles \cite{5} whereas lactones and esters were found to be scarce in the flavor of specific cultivars \cite{5}. In this work, we  selected four prominent volatiles, that are emitted in a high amount from mango during the ripening periods and studied the odor quality of mango fruit with the support of Gdn-NS. The interaction behavior of ethyl butanoate, myrcene, (E,Z,Z)-1,3,4,8-undecatetraene and $\gamma$-octalactone aromas from ripening periods of mango fruit has been studied upon adsorption of Gdn-NS. Figure~\ref{fig-s4} represents the adsorption of ethyl butanoate, myrcene, (E,Z,Z)-1,3,4,8-undecatetraene and $\gamma$-octalactone on Gdn-NS. All the positions of volatiles emitted from the mango fruit adsorbed on Gdn-NS are global minima positions. The interaction of O-atom in ethyl butanoate and $\gamma$-octalactone on C atom in Gdn-NS is represented as 1-Gdn and 4-Gdn, respectively. In the case of corresponding interaction of hydrogen atom in myrcene and (E,Z,Z)-1,3,4,8-undecatetraene on C atom in Gdn-NS is referred to as 2-Gdn and 3-Gdn. In this proposed work, calculating the adsorption energy, Bader charge transfer, the average band gap changes upon adsorption of volatiles from the mango fruit with Gdn-NS are utilized to examine the odor quality of the mango fruit. The negative adsorption energy ($E_{\text{ad}}$) is detected upon the interaction of ethyl butanoate, myrcene, (E,Z,Z)-1,3,4,8-undecatetraene and $\gamma$-octalactone on Gdn-NS, which clearly demonstrates the adsorption of Mangifera indica L. volatiles on graphdiyne nanosheet. 

The $E_{\text{ad}}$ of Gdn-NS for the corresponding adsorption sites from 1-Gdn to 4-Gdn are found to be $-0.280$, $-0.520$, $-1.012$ and $-0.247$ eV. Nevertheless, the energy band gap of graphdiyne nanosheet gets changed due to the interaction of such aromas in Gdn-NS. During the adsorption of the mango volatiles on Gdn-NS, it leads to transmission of electrons between VOCs and Gdn-NS material that results in  the changes in Gdn-NS electrical resistance. The changes in Gdn-NS conductance or resistance can be examined with the two-probe device \cite{54}. Further, to investigate the influence of thermodynamic property on the interaction behavior of aforementioned mango VOCs on Gdn-NS material, Gibbs-free-energy changes $(\Delta G)$ are studied at a usual temperature and standard pressure (298 K, 1 atmosphere) using the expression below

\begin{equation}
\Delta G_{\text{ad}} = [G(\text{Gdn+VOC})-G(\text{Gdn})-G(\text{VOC})].
\label{eq3.2}
\end{equation}

	The variation in $\Delta G$ energies is comparatively lower than their corresponding value of adsorption energy due to the entropy and temperature effects as shown in table~\ref{table 1}. As a result from frequency analysis, the magnitudes of $\Delta G$ for the adsorption sites from 1-Gdn to 4-Gdn are noticed to be $-0.223$, $-0.456$, $-0.947$ and $-0.188$~eV, respectively. Importantly, the selection of exchange-correlation functional highly influences $\Delta G$. The $\Delta G$ of the aforementioned sites with various exchange-correlation functional is displayed in table~\ref{table 2}. It is revealed that the selection of hybrid functional leads to the increase of $\Delta G$. Further, the deviation in the $E_g$ of Gdn-NS upon interaction of such volatiles results in the deviation in the conductivity of Gdn base material. Precisely, the changes in $E_g$ of Gdn-NS for the respective positions from 1-Gdn to 4-Gdn is recorded to be 0.12, 0.38, 0.32 and 0.54 eV. Supportively, the changes in the adsorption energy and energy gap were perceived upon the interaction of aromas from ripening stage of~mangoes.

\begin {table}[!t]
  \renewcommand{\arraystretch} {1.3}
  \caption{ Significant parameters for adsorption studies of Gdn-NS with mango volatiles.}
  \label{table 1}
  \vspace{3mm}
  \centering
  \scriptsize\begin{tabular}{|c|c|c|c|c|c|}
  \hline
  Graphdiyne Nanosheet  	&Adsorption energy  
	& Gibbs free energy change 
 &Bader charge transfer  &Energy gap  &$E_g^a$\,, \% \\
	
	and positions &$E_{\text{ad}}$ (eV) &$\Delta G$ – Graphdiyne nanosheet (eV) &Q (e) &$E_g$ (eV) &\\
	
	\hline
	
Gdn nanosheet	&-	&-	&- &0.51	&-\\

\hline
Position 1-Gdn	&$-0.280$	& $-0.223$ &0.454	&0.12	&76.47 \\

\hline
Position 2-Gdn	&$-0.520$	& $-0.456$ &0.129	&0.38	&25.49 \\
\hline
Position 3-Gdn	&$-1.012$	& $-0.947$ &0.072	&0.32	&37.25 \\
\hline
Position 4-Gdn	&$-0.247$	& $-0.188$ &0.465	&0.54	&5.88 \\
\hline

  \end {tabular}
  \end{table}
	\normalsize

	\begin {table}[!t]
  
  \renewcommand{\arraystretch} {1.3}
  \caption{ Comparative statement of Gibbs free energy change $\Delta G$ with different exchange-correlation functional.}
  \label{table 2}
   \vspace{3mm}
  \centering
  \scriptsize\begin{tabular}{|c|c|c|c|c|}
  \hline
  \multicolumn{5}{|c|}{Gibbs free energy change $\Delta G$ – Graphdiyne nanosheet (eV)} \\
	\hline
	 Positions &PBE &BLYP  &revPBE  &RPBE \\
	
	\hline
	
Position 1-Gdn	&$-0.223$	&$-0.390$	&$-0.326$ &$-0.390$\\

\hline
Position 2-Gdn	&$-0.456$	& $-0.537$ &$-0.514$	&$-0.518$	 \\
\hline
Position 3-Gdn	&$-0.947$	& $-1.132$ &$-1.033$	&$-1.042$	 \\
\hline
Position 4-Gdn	&$-0.188$	& $-0.239$ &$-0.219$	&$-0.239$	 \\
\hline

  \end {tabular}
  \end{table}
	\normalsize

\begin{figure}[!b]
\centerline{\includegraphics[width=0.85\textwidth]{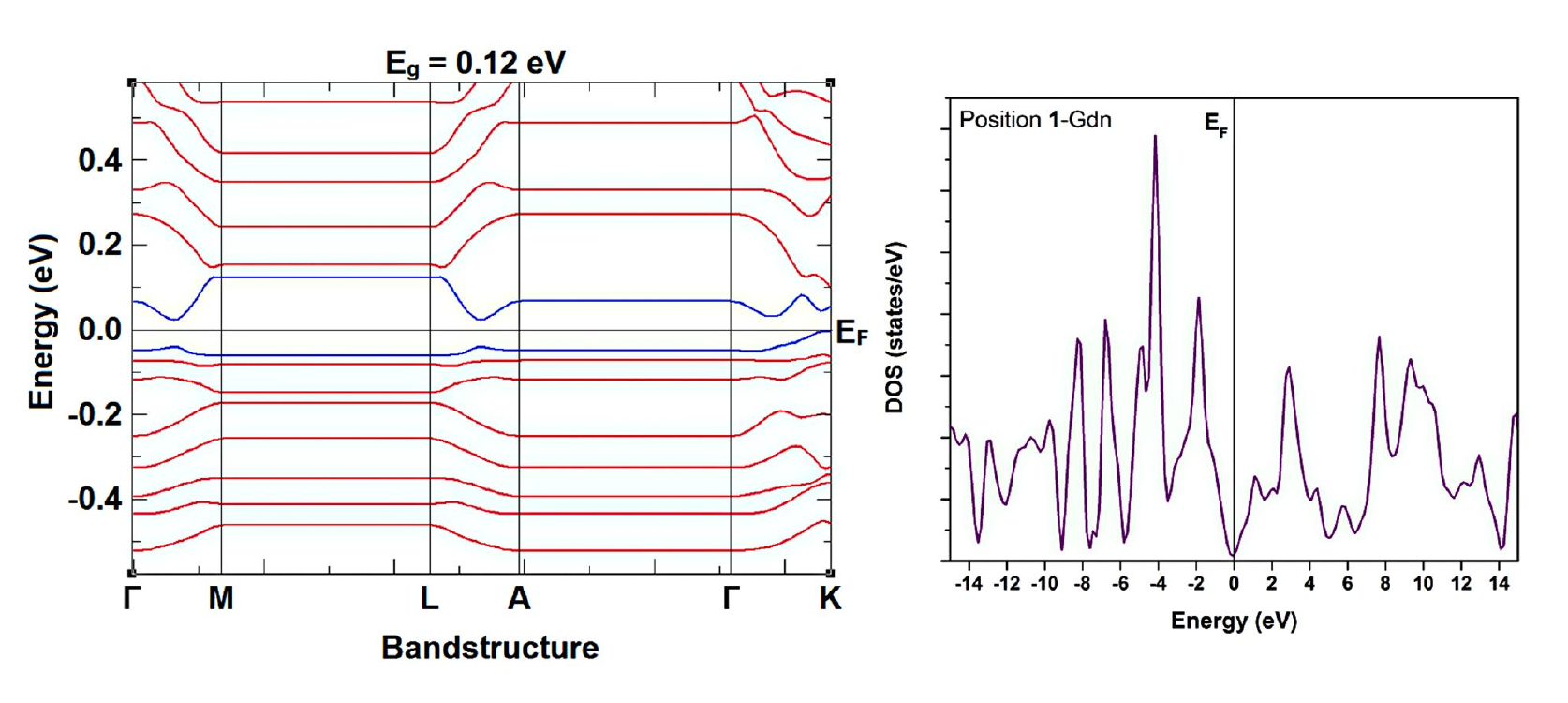}}
\caption{(Colour online) Band structure and DOS spectrum – position 1-Gdn.} \label{fig-s5}
\end{figure}

\begin{figure}[!b]
\centerline{\includegraphics[width=0.85\textwidth]{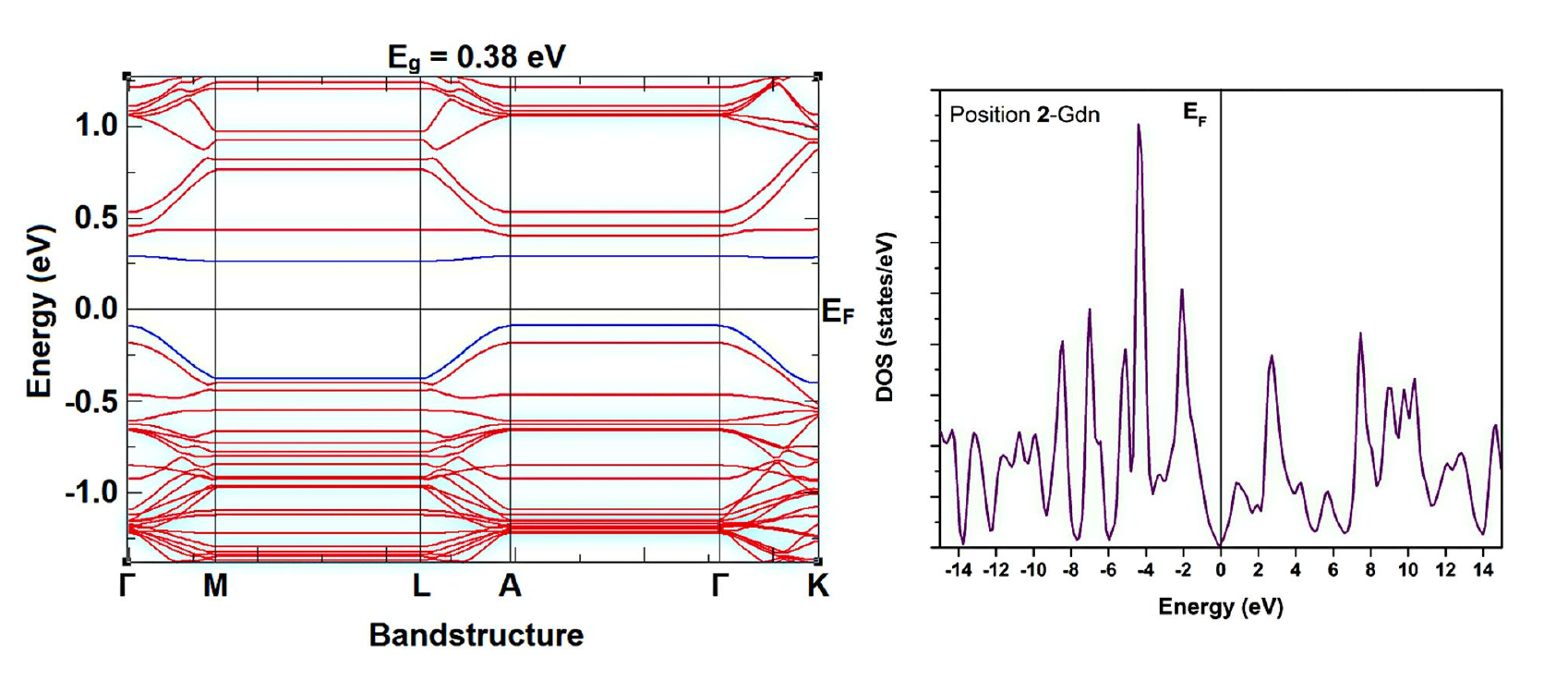}}
\caption{(Colour online) Band structure and DOS spectrum – position 2-Gdn.} \label{fig-s6}
\end{figure}

\begin{figure}[!b]
\centerline{\includegraphics[width=0.85\textwidth]{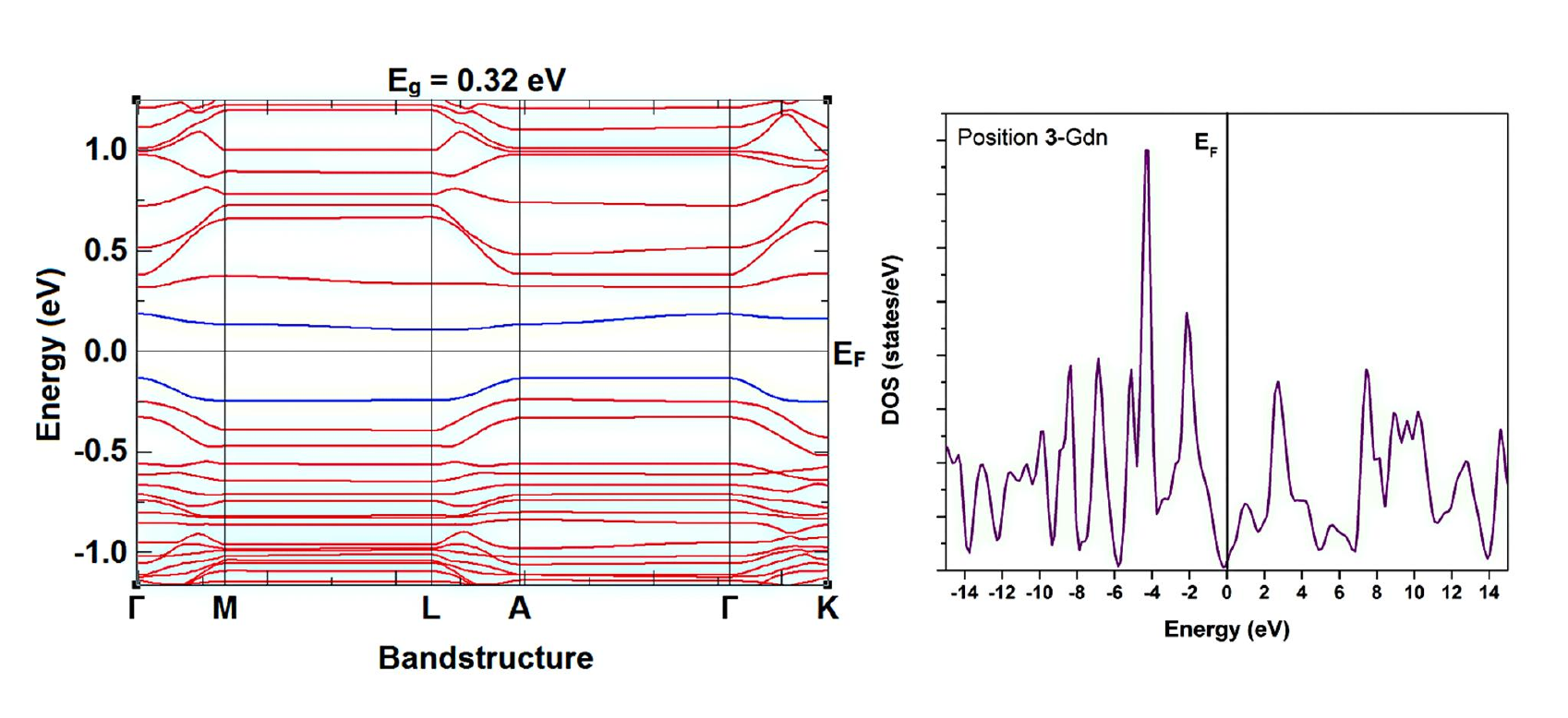}}
\caption{(Colour online) Band structure and DOS spectrum – position 3-Gdn.} \label{fig-s7}
\end{figure}

\begin{figure}[htb]
\centerline{\includegraphics[width=0.85\textwidth]{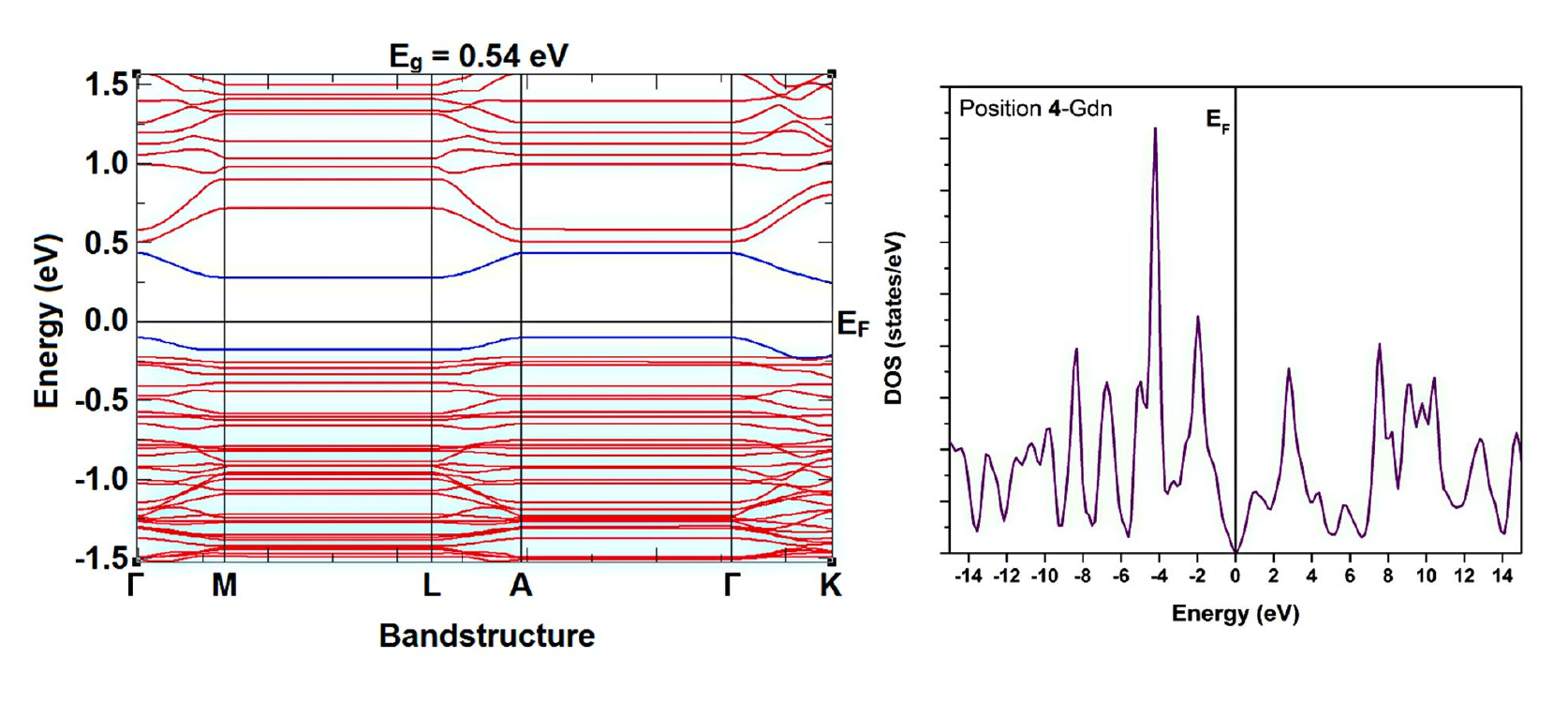}}
\caption{(Colour online) Band structure and DOS spectrum – position 4-Gdn.} \label{fig-s8}
\end{figure}

		Many authors have recommended  the Gdn-NS as a promising material for the probing of various toxic vapor and gas molecules including H$_2$, CH$_4$, and CO \cite{29}. Chen and co-workers \cite{30}  reported the sensing behavior of amino acids on Gdn base material. We have already reported the interaction behavior of trimethyl amine, NH$_3$ and dimethyl amine on Gdn-NS using first-principle calculations. Furthermore, numerous reports suggested that the quality of the mango fruit can be determined using non-destructive methods through electronic noses \cite{7,8,9,10}. Till now, up to our knowledge, no work has been carried out based on monolayer Gdn-NS to check the odor quality of mango using DFT method. For the very first time, apparently we are demonstrating the work for odor quality checking of the mango fruit using ab initio calculation. Notably, the most deciding and significant factor for probing the odor quality of mango is average-band-gap changes ($E_g^a$) compared with pristine Gdn-NS to Gdn-NS/VOCs complex. Table~\ref{table 1} contains the information regarding the average-band-gap changes, adsorption energy, energy gap and Bader charge transfer, of Gdn-NS upon interaction of mango volatiles. Besides, another question may arise. How to differentiate the odor of volatiles emitted from the ripening stage of mango using Gdn nanosheet? As a result obtained from various parameters of adsorption study, it is precisely confirmed that the interaction of aroma volatiles from the ripening stages of mango, namely ethyl butanoate, myrcene, (E,Z,Z)-1,3,4,8-undecatetraene and $\gamma$-octalactone on Gdn-NS shows various sensing response supported by $E_g^a$. It is inferred that based on different sensing response shown by Gdn-NS, we can quickly identify the odor quality of mango fruit. Therefore, the deviation in the energy gap of Gdn-NS upon interaction of these volatiles is directly proportional to the odor quality of the mango fruit. For instance, position 1-Gdn shows the maximum sensing response ($E_g^a$ = 76.47\%) towards ethyl butanoate, and the corresponding odor of this volatile is fruity as we know from the literature reports~\cite{55}. Similarly, the corresponding sensing response for the positions 2-Gdn (25.49\%), 3-Gdn (37.25\%) and 4-Gdn (5.88\%) towards myrcene, (E,Z,Z)-1,3,4,8-undecatetraene and  $\gamma$-octalactone confirms the odor quality of the mango to be terpeny~\cite{56}, pineapple \cite{57} and coconut~\cite{56}. Figures~\ref{fig-s5}, \ref{fig-s6}, \ref{fig-s7}, \ref{fig-s8} refers to the pictographic representation of the band structure and DOS-spectrum for the positions from 1-Gdn to 4-Gdn, respectively. From the result of DOS-spectrum and band structure, more $P_{\text{MAX}}$ are detected on the conduction band (lowest unoccupied – molecular orbitals (LUMO)) of Gdn-NS. The DOS-spectrum of Gd-NS validates the interaction of the mango VOCs on Gdn nanosheet. For instance, during the transfer of negatively charged electrons between Gdn-NS and mango VOCs, the charge density differs through various energy intervals of Gdn material especially in LUMO level. Besides, high sensing response is noticed upon adsorption of ethyl butanoate on Gdn nanosheet. It infers that the stage of mango fruit is shifting to the overripe stage and reveals that the odor quality is fruity. 
	
\begin{figure}[!b]
\centerline{\includegraphics[width=0.65\textwidth]{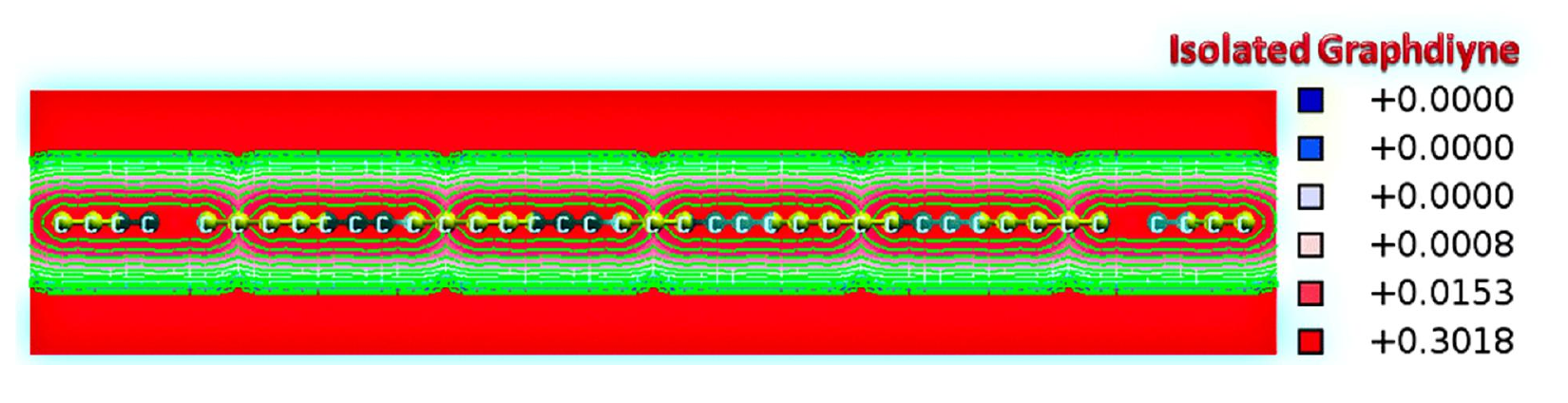}}
\caption{(Colour online) Electron density of isolated Gdn-NS.} \label{fig-s9}
\end{figure}

\begin{figure}[!b]
\centerline{\includegraphics[width=0.65\textwidth]{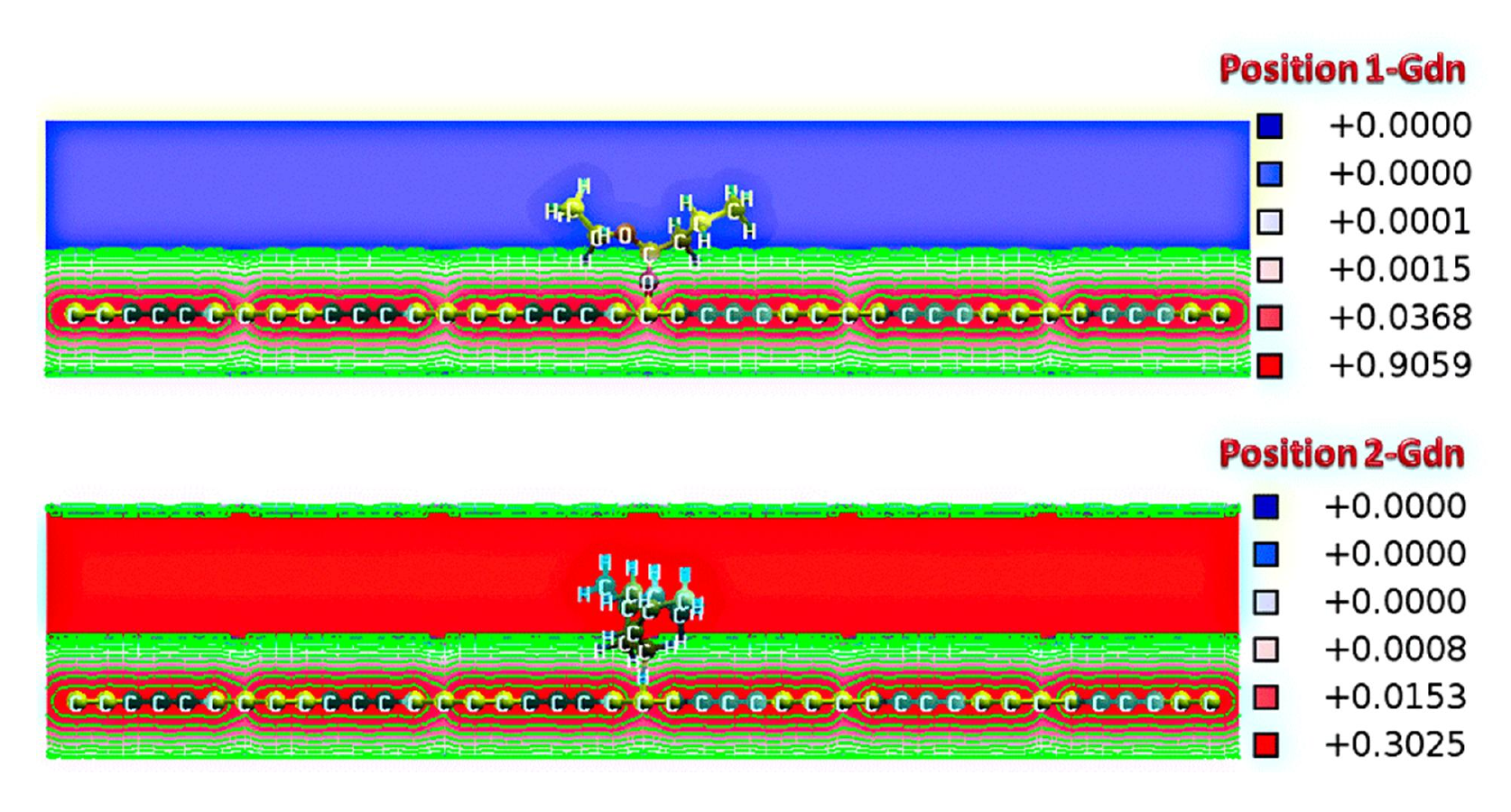}}
\caption{(Colour online) Electron density – Position 1-Gdn \& 2-Gdn.} \label{fig-s10}
\end{figure}

\begin{figure}[!t]
\centerline{\includegraphics[width=0.65\textwidth]{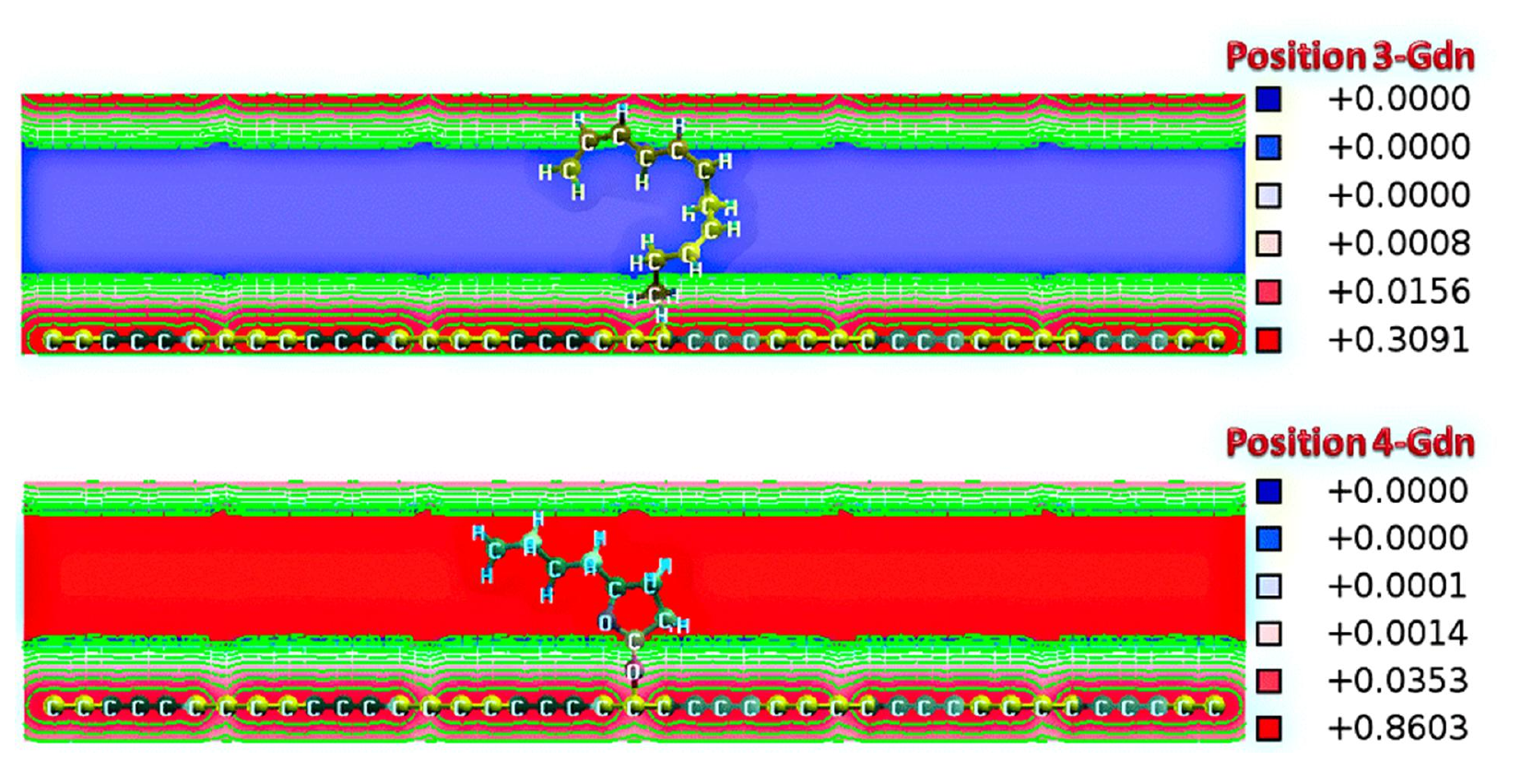}}
\caption{(Colour online) Electron density – Position 3-Gdn \& 4-Gdn.} \label{fig-s11}
\end{figure}

The electron density portrait (ED) of Gdn-NS upon interaction of the aforementioned volatiles is shown in figures~\ref{fig-s9}, \ref{fig-s10}, \ref{fig-s11}. From the detailed pictorial representation of ED, it is verified that the density of electrons is detected to be higher beside the interaction sites with Gdn-NS. Further, to succeed the prime focus of the present work, (i.e.,) to detect the odor quality of the mango fruit, the sensitivity of Gdn-NS towards ethyl butanoate, myrcene, (E,Z,Z)-1,3,4,8-undecatetraene and $\gamma$-octalactone aromas must be studied. The DOS-spectrum analysis of all global minima adsorption sites on Gdn-NS must be compared with their corresponding pristine Gdn-NS material.

The conductivity $(\sigma)$ of 2D-Gdn material can be studied using the  traditional equation~(\ref{eq3.3}):

\begin{equation}
\sigma = \text{A} \exp (-E_g/2kT),
\label{eq3.3}
\end{equation}
\begin{figure}[!b]
\centerline{\includegraphics[width=0.65\textwidth]{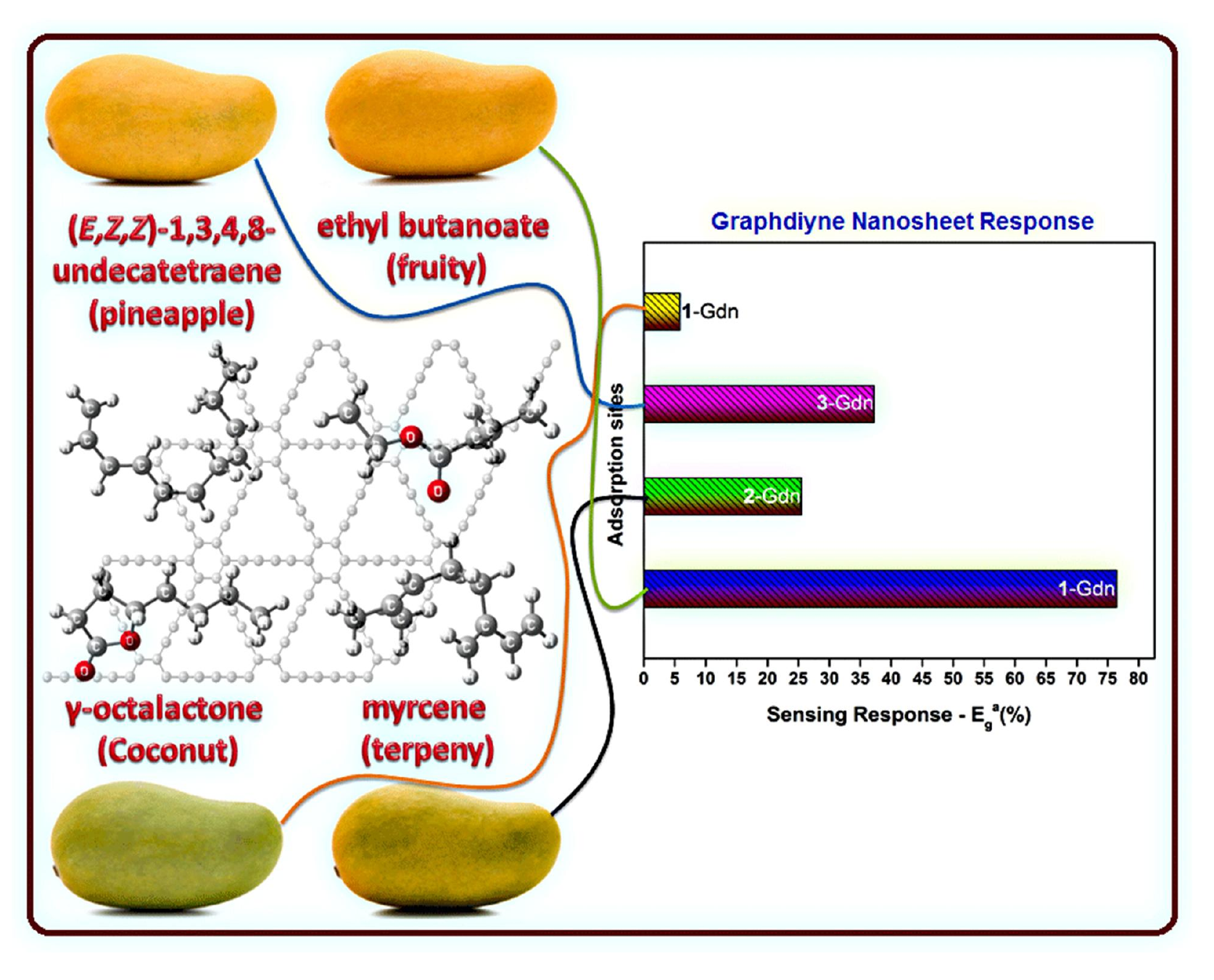}}
\caption{(Colour online) The discernments on the odor quality of mango from various ripening stages using Gdn-NS monolayers.} \label{fig-s12}
\end{figure}
where $k$ and $\sigma$ represents the Boltzmann’s constant and electrical conductivity, respectively. ‘A’ implies the proportionality constant for Gdn-NS material. Based on the overhead traditional expression, it is well known that low magnitude of $E_g$ leads to high electrical conductivity. Further, for all the possible interaction sites from 1-Gdn to 4-Gdn, the $E_g$ deviates from 0.51 for pristine Gdn-NS to 0.12, 0.38, 0.32 and 0.54 eV for the corresponding adsorption sites. Comparatively, more variation is noticed upon the interaction of ethyl butanoate, which confirms that the fruit is in overripened stage \cite{58,59}. In the case of positions 2-Gdn, 3-Gdn and 4-Gdn, not much deviation is found upon the interaction of myrcene, (E,Z,Z)-1,3,4,8-undecatetraene and  $\gamma$-octalactone on Gdn-NS, which means that the mango fruits are in ripening stage. Therefore, based on our previous work, we have reported the ripe and overripe stage of fruit through the percentage of variation in $E_g$ upon the interaction of mango volatiles on Gdn-NS material.
	The other most important and deciding factor for exploring the interaction of ethyl butanoate, myrcene, (E,Z,Z)-1,3,4,8-undecatetraene and  $\gamma$-octalactone aromas in Gdn-NS is Bader atoms in molecules (AIM) analysis~(Q). The objective of calculating the Bader charge transfer (BCT) is to determine the transfer of free electrons among Gdn-NS and mango aromas \cite{40,60,61,62,63,64}.  The BCT for the interaction sites from 1-Gdn to 4-Gdn are noticed to be 0.454$e$, 0.129$e$, 0.072$e$ and 0.465$e$, respectively. The transfer of charge amongst Gdn-NS and aforementioned VOCs is observed with the positive magnitude of Q. This confirms that the free electrons migrate from Mangifera indica L.aromas to Gdn nanosheet. However, the magnitude of the positions 1-Gdn and 4-Gdn is almost the same, high sensing response is observed only in 4-Gdn, which arose due to the overripened stage of the mango fruit. From the overall results, it is revealed that the ripe (overripe) stage and odor quality of mango can be easily identified with the support of average energy gap changes (sensing response). Figure~\ref{fig-s12} refers to the perception on the odor quality of mango from various ripening stages using Gdn-NS monolayers. Therefore, Gdn nanosheet is highly sensitive towards mango volatiles, when the fruits are in over-ripened stage, and weakly sensitive to the ripened stage of mango VOCs. Thus, the odor quality, as well as the ripening stage of mango fruit, can be indeed checked using Gdn nanosheet.

\section {Conclusion}

With ab initio calculation, we explored the geometric stability of Gdn-NS using phonon band gap structure and cohesive formation energy. In this work, graphdiyne base material was used to check the odor quality of Mangifera indica L. fruit. The adsorption of different volatiles namely: (1) ethyl butanoate, (2) myrcene, (3) (E,Z,Z)-1,3,4,8-undecatetraene and (4) $\gamma$-octalactone from various ripening stages of mango on Gdn-NS is investigated from the adsorption energy, energy gap, average energy gap changes and Bader charge transfer. Precisely, the sensitivity of such aromas interacted on Gdn-NS is also validated with the DOS-spectrum. Comparatively, a higher sensing response has been noticed upon interaction of ethyl butanoate (fruity odor) rather than myrcene (terpeny), (E,Z,Z)-1,3,4,8-undecatetraene (pineapple) and  $\gamma$-octalactone (coconut). The results suggest that the Gdn nanosheet is an appropriate base material to check the odor quality of the mango fruit.

\section {Acknowledgements}

The authors wish to express their sincere thanks to Nano Mission Council (No.~SR/NM/NS-1011/2017(G)) Department of Science \& Technology, India for  financial support.
\section {Data availability}
The raw/processed data required to reproduce these findings cannot be shared at this time because the data also form part of an ongoing study.

\ukrainianpart

\title
{Виявлення якості запаху та стану достигання плодів манго за допомогою графдійнового нанолиста --- погляд DFT}%

\author{В. Нагараян, Р. Чандірамулі}
\address{
Школа електротехнiки та електроніки, унiверситет SASTRA, Тiрумалайсамудрам, Танджавур	
--- 613 401, Індія 
}

\makeukrtitle

\begin{abstract}
Використовуючи першопринципні обчислення, досліджено геометричну стійкість разом з електронними властивостями графдійного нанолиста (Gdn-NS). Структурна стійкість  Gdn-NS встановлюється з допомогою фононної зонної структури і когезійної енергії. Основною метою даного   дослідження є перевірка якості запаху плодів манго Mangifera indica L. протягом  різних етапів дозрівання під дією матеріалу  Gdn-NS. Окрім цього, досліджується адсорбція різних летких речовин, зокрема, ароматів етил бутанату, мірцену,
 (E,Z,Z)-1,3,4,8-андекатетрену та  $\gamma$-окталактону на Gdn-NS разом з визначенням таких важливих параметрів як перенесення заряду Бейдера, ширина забороненої зони, зміна середньої ширини забороненої зони та енергія поглинання. Чутливість речовин, випромінюваних на різних етапах дозрівання манго, досліджувалося під дією спектру густини станів. Результати даної роботи допомогли перевірити  етап дозрівання  та якість запаху плодів  манго  Mangifera indica L. з допомогою  Gdn-NS матеріалу, використовуючи теорію функціоналу густини.
\keywords графдійн, адсорбція, нанолист, ширина забороненої зони, когезійна енергія  %

\end{abstract}
\lastpage

\begin{thebibliography}{99}
\bibitem{1}	Chauhan O.P., Raju P.S., Bawa A.S., Mango flavor. In: Handbook of Fruit and Vegetable Flavors, Hui, Y.H., (Ed.), Wiley, Hoboken, NJ, USA, 2010, 319--343.
\bibitem{2}	VCF Volatile Compounds in Food: Database, version 14.1, Nijssen L.M., Ingen-Visscher C.A.V., \\ Donders~J.J.H., (Eds.), TNO Triskelion: Zeist, Netherlands, 2014, 1963--2013,  URL~
\url{http://www.vcf-online.nl.}
\bibitem{3}	Kulkarni R.S., Chidley H.G., Pujari K.H., Giri A.P., Gupta V.S., Food Chem., 2012, \textbf{130}, 58,\\ \doi{10.1016/j.foodchem.2011.06.053}.
\bibitem{4}	Schieberle P., Grosch W.,  Z. Lebensm  Unters. Forsch., 1987, \textbf{185}, 111, \doi{10.1007/BF01850088}.
\bibitem{5}	Lopes D.C., Fraga S.R., Rezende C.M., Quim. Nova, 1999, \textbf{22}, 31, \\ \doi{10.1590/S0100-40421999000100007}.	
\bibitem{6}	Pino J.A., Mesa J., Flavour Fragrance J., 2006, \textbf{21}, 207, \doi{10.1002/ffj.1703}.
\bibitem{7}	Jha S.N., Narsaiah K., Sharma A.D., Singh M., Bansal S., Kumar R., J. Food Sci. Technol., 2010, \textbf{47}, 1,\\ \doi{10.1007/s13197-010-0004-6}.
\bibitem{8}	Rock F., Barsan N., Weimar U., Chem. Rev., 2008, \textbf{108}, 705, \doi{10.1021/cr068121q}.
\bibitem{9}	Saranwong S., Sornsrivichai J., Kawano S., Postharvest Biol. Technol., 2004, \textbf{31}, 137, \\ \doi{10.1016/j.postharvbio.2003.08.007}.
\bibitem{10}	Lebrun M., Plotto A., Goodner K., Ducamp M.-N., Baldwin E., Postharvest Biol. Technol., 2008, \textbf{48}, 122,\\ \doi{10.1016/j.postharvbio.2007.09.010}.
\bibitem{11}	Pawliszyn J., Solid Phase Microextraction: Theory and Practice, Wiley-VCH Inc., New York, 1997.
\bibitem{12}	Lindinger W., Hansel A., Jordan A., Int. J. Mass Spectrom., 1998, \textbf{173}, 191,  \\ \doi{10.1016/S0168-1176(97)00281-4}.
\bibitem{13}   Novoselov K.S., Fal'ko V.I., Colombo L., Gellert P.R., Schwab M.G., Kim K., Nature, 2012, \textbf{490}, 192, \\ \doi{10.1038/nature11458}.   
\bibitem{14}		Kroto H.W., Heath J.R., O'Brien S.C., Curl R.F., Smalley R.E., Nature, 1985, \textbf{318}, 162, \\ \doi{10.1038/318162a0}.
\bibitem{15}		Kong X.Y., Ding Y., Yang R., Wang Z.L., Science, 2004, \textbf{303}, 1348, \doi{10.1126/science.1092356}.
\bibitem{16}		Nasibulin A.G., Pikhitsa P.V., Jiang H., Brown D.P., Krasheninnikov A.V., Anisimov A.S., Queipo P., Moisala~A., Gonzalez D., Lientschnig G., Hassanien A., Shandakov S.D., Lolli G., Resasco D.E., Choi M., Tom\'anek D., Kauppinen E.I.,  Nat. Nanotechnol., 2007, \textbf{2}, 156, \doi{10.1038/nnano.2007.37}.
\bibitem{17}		Hirsch A., Nat. Mater., 2010, \textbf{9}, 868, \doi{10.1038/nmat2885}.
\bibitem{18}		Kang J., Li J., Wu F., Li S.S., Xia J.B., J. Phys. Chem. C, 2011, \textbf{115}, 20466, \doi{10.1021/jp206751m}.
\bibitem{19}		Srinivasu K., Ghosh S.K., J. Phys. Chem. C, 2012, \textbf{116}, 5951, \doi{10.1021/jp212181h}.
\bibitem{20}		Ivanovskii A.L., Prog. Solid State Chem., 2013, \textbf{41}, 1, \doi{10.1016/j.progsolidstchem.2012.12.001}.
\bibitem{21}		Pei Y., Physica B, 2012, \textbf{407}, 4436, \doi{10.1016/j.physb.2012.07.026}.
\bibitem{22}	Haley M.M., Brand S.C., Pak J.J., Angew. Chem. Int. Ed. Eng., 1997, \textbf{36}, 836, \\\doi{10.1002/anie.199708361}.
\bibitem{23}	Li G., Li Y., Liu H., Guo Y., Li Y., Zhu D., Chem. Commun., 2010, \textbf{46}, 3256, \doi{10.1039/b922733d}.
\bibitem{24}	Haley M.M., Pure Appl. Chem., 2008, \textbf{80}, 519, \doi{10.1351/pac200880030519}.
\bibitem{25}	Yang N.L., Liu Y.Y., Wen H., Tang Z.Y., Zhao H.J., Li Y.L., Wang D., ACS Nano, 2013, \textbf{7}, 1504,\\ \doi{10.1021/nn305288z}.
\bibitem{26}	He J.J., Ma S.Y., Zhou P., Zhang C.X., He C.Y., Sun L.Z., J. Phys. Chem. C, 2012, \textbf{116}, 26313,\\ \doi{10.1021/jp307408u}.
\bibitem{27}	Wu P., Du P., Zhang H., Cai C.X., Phys. Chem. Chem. Phys., 2014, \textbf{16}, 5640.
\bibitem{28}	Pan Y.Y., Wang Y.Y., Wang L., Zhong H., Quhe R., Ni Z., Ye M., Mei W.-N., Shi J.J., Guo W., \\Yang J., Lu J., Nano, 2015, 
				\textbf{7}, 2116.
\bibitem{29}	Jiao Y., Du A.J., Hankel M., Zhu Z.H., Rudolph V., Smith S.C., Chem. Commun., 2011, \textbf{47}, 11843,\\ \doi{10.1039/c1cc15129k}.
\bibitem{30}	Chen X., Gao P., Guo L., Zhang S., Sci. Rep., 2015, \textbf{5}, 16720, \doi{10.1038/srep16720}.
\bibitem{31}	Nagarajan V., Srimathi U., Chandiramouli R., Comput. Theor. Chem., 2018, \textbf{1123}, 119, \\ \doi{10.1016/j.comptc.2017.11.020}.
\bibitem{32}	Nagarajan V., Chandiramouli R., J. Mol. Liq., 2018, \textbf{249}, 24, \doi{10.1016/j.molliq.2017.11.007}.
\bibitem{33}	Soler J.M., Artacho E., Gale J.D., Garcia A., Junquera J., Ordejon P., S-Portal D., J. Phys. \\Condens. Matter, 2002, \textbf{14}, 2745, \doi{10.1088/0953-8984/14/11/302}.
\bibitem{34}	Roman-Perez G., Soler J.M., Phys. Rev. Lett., 2009, \textbf{103}, 096102, \doi{10.1103/PhysRevLett.103.096102}.
\bibitem{35}	Luo G., Zheng Q., Mei W.-N., Lu J., Nagase S., J. Phys. Chem. C, 2013, \textbf{117}, 13072, \doi{10.1021/jp402218k}.
\bibitem{36}	Troullier N., Martins J., Solid State Commun., 1990, \textbf{74}, 613, \doi{10.1016/0038-1098(90)90686-6}.
\bibitem{37}	Princy Maria J., Nagarajan V., Chandiramouli R., Chem. Phys. Lett., 2020, \textbf{738}, 136841,\\ \doi{10.1016/j.cplett.2019.136841}.
\bibitem{38}	Bhuvaneswari R., Nagarajan V., Chandiramouli R., Chem. Phys., 2020, \textbf{530}, 110604,\\ \doi{10.1016/j.chemphys.2019.110604}. 
\bibitem{39}	Monkhorst H.J., Pack J.D., Phys. Rev. B, 1976, \textbf{13}, 5118, \doi{10.1103/PhysRevB.13.5188}. 
\bibitem{40}	Bader R., Atoms in Molecules: A Quantum Theory, Oxford University Press, New York, 1990.
\bibitem{41}	Kim J., Esler K.P., McMinis J., Morales M.A., Clark B.K., Shulenburger L., Ceperley D.M., J. Phys.: Conf. Ser., 2012, \textbf{402}, 012008, \doi{10.1088/1742-6596/402/1/012008}. 
\bibitem{42}	Hood R.Q., Kent P.R.C., Reboredo F.A., Phys. Rev. B, 2012, \textbf{85}, 134109, \doi{10.1103/PhysRevB.85.134109}.
\bibitem{43}	Benali A., Shulenburger L., Krogel J.T., Zhong X., Kente P.R.C., Heinonen O., Phys. Chem. Chem. Phys., 2016, \textbf{18}, 18323, \doi{10.1039/C6CP02067D}.
\bibitem{44}	Swetha B., Nagarajan V., Chandiramouli R., ChemistrySelect, 2019, \textbf{4}, 14237, \doi{10.1002/slct.201903737}
\bibitem{45}	Beheshtian J., Peyghan A.A., Noei M., Sens. Actuators, B, 2013, \textbf{181}, 829, \doi{10.1016/j.snb.2013.02.086}.
\bibitem{46}	Beheshtian J., Peyghan A.A., Bagheri Z., Physica E, 2012, \textbf{44}, 1963, \doi{10.1016/j.physe.2012.06.003}.
\bibitem{47}	Long M., Tang L., Wang D., Li Y., Shuai Z., ACS Nano, 2011, \textbf{5}, 2593, \doi{10.1021/nn102472s}.
\bibitem{48}	Shin H., Kang S., Koo J., Lee H., Kim J., Kwon Y., J. Chem. Phys., 2014, \textbf{140}, 114702, \doi{10.1063/1.4862829}.
\bibitem{49}	Shin H., Kim J., Lee H., Heinonen O., Benali A., Kwon Y., J. Chem. Theory Comput., 2017, \textbf{13}, No.~11, 5639, \doi{10.1021/acs.jctc.7b00747}.
\bibitem{50}	Nulakani N.V.R., Subramanian V., ACS Omega, 2017, \textbf{2}, 6822, \doi{10.1021/acsomega.7b00513}.
\bibitem{51}	Bhuvaneswari R., Nagarajan V., Chandiramouli R., Molecular Physics, 2019,\\ \doi{10.1080/00268976.2019.1699184}.
\bibitem{52}	Zhang S., Hu Y., Hu Z., Cai B., Zeng H., Appl. Phys. Lett., 2015, \textbf{107}, 022102, \doi{10.1063/1.4926761}.
\bibitem{53}	Nijssen I.M., Visscher C.A., Maarse H., Willemsens I.C., Boelens M.H., TNO Nutrition and Food Research Institute: Zeist, 1996, 1–10.
\bibitem{54}	Prasongkit J., Amorim R.G., Chakraborty S., Ahuja R., Scheicher R.H., Amornkitbamrung V., J. Phys. Chem.~C, 2015, \textbf{119}, 16934, \doi{10.1021/acs.jpcc.5b03635}.
\bibitem{55}	Engel K.H., Tressl R., J. Agric. Food Chem., 1983, \textbf{31}, 796, \doi{10.1021/jf00118a029}.
\bibitem{56}	Hunter G.L.K., Bucek W.A., Radford T., J. Food Sci., 1974, \textbf{39}, 900, \doi{10.1111/j.1365-2621.1974.tb07271.x}.
\bibitem{57}	Berger R.G., Drawert F., Kollmannsberger H., Nitz S., J. Food Sci., 1985, \textbf{50},  1655, \\\doi{10.1111/j.1365-2621.1985.tb10558.x}.
\bibitem{58}	Nagarajan V., Chandiramouli R., Condens. Matter Phys., 2019, \textbf{22}, 13703, \doi{10.5488/CMP.22.13703}.
\bibitem{59}	Keerthi Bhavadharani R., Nagarajan V., Chandiramouli R., Condens. Matter Phys., 2019, \textbf{22}, 33001, \\ \doi{10.5488/CMP.22.33001}.
\bibitem{60}	Princy Maria J., Bhuvaneswari R., Nagarajan V., Chandiramouli R., J. Mol. Graphics Modell., 2020, \textbf{95}, 107505, \doi{10.1016/j.jmgm.2019.107505}.
\bibitem{61}	Zhang S., Guo S., Chen Z., Wang Y., Gao H., Herrero J. G., Ares P., Zamora F., Zhu Z., Zeng H., Chem. Soc. Rev., 2018, \textbf{47}, 982, \doi{10.1039/C7CS00125H}.
\bibitem{62}	Ullah H., Shah A. H. A., Bilal S., Ayub K., J. Phys. Chem. C, 2014, \textbf{118}, 17819, \doi{10.1021/jp505626d}.
\bibitem{63}	Chigo Anota E., Cortes Arriagada D., Bautista Hern\'andez A., Castro M., Appl. Surf. Sci., 2017, \textbf{400}, 283, \doi{10.1016/j.apsusc.2016.12.153}.
\bibitem{64}	Bhuvaneswari R., Nagarajan V., Chandiramouli R., Phys. Lett. A, 2019, \textbf{383}, 125975,\\ \doi{10.1016/j.physleta.2019.125975}.
\end{thebibliography}
\end{document}